\documentclass[conference]{IEEEtran}
\IEEEoverridecommandlockouts
\usepackage{cite}
\usepackage{amsmath,amssymb,amsfonts}
\usepackage{algorithmic}
\usepackage{graphicx}
\usepackage{textcomp}
\usepackage{xcolor}
\usepackage{subcaption}
\usepackage{listings}
\usepackage{enumitem}
\usepackage[ruled]{algorithm2e}
\usepackage{caption}
\DeclareCaptionFont{8pt}{\fontsize{8pt}{6pt}\selectfont}
\def\BibTeX{{\rm B\kern-.05em{\sc i\kern-.025em b}\kern-.08em
    T\kern-.1667em\lower.7ex\hbox{E}\kern-.125emX}}
\begin{document}

\makeatletter 
\newcommand{\linebreakand}{%
  \end{@IEEEauthorhalign}
  \hfill\mbox{}\par
  \mbox{}\hfill\begin{@IEEEauthorhalign}
}
\makeatother 

\title{DAOS as HPC Storage: Exploring Interfaces}

\author{
\IEEEauthorblockN{Adrian Jackson}
\IEEEauthorblockA{\textit{The University of Edinburgh, United Kingdom} \\
a.jackson@epcc.ed.ac.uk}
\and
\IEEEauthorblockN{Nicolau Manubens}
\IEEEauthorblockA{\textit{European Centre for Medium-Range} \\
\textit{Weather Forecasts, Germany} \\
nicolau.manubens@ecmwf.int}
}

\maketitle

\section{Introduction}

Reading and writing data to and from data storage has long been a bottleneck for high performance computing (HPC). The hardware used for data storage exhibits performance that is generally at least an order of magnitude lower than that of volatile memory or processing hardware.  Applications often limit the amount of I/O undertaken on HPC systems to reduce this cost, but recently there has been a rise in application categories where ingestion or production of large amounts of data is common, machine learning being an obvious example. As HPC systems increase in size, reaching Exascale levels and beyond, and a subset of applications require ever larger amounts of I/O bandwidth or metadata performance, there is a significant challenge to improve the performance of the data storage technologies employed for I/O operations.

Object storage technology has gained significant traction in the web services and cloud markets for data that does not necessarily fit into standard filesystem structures or require more storage than databases traditionally provide. However, even the object stores that are used in scientific computing, such as Ceph~\cite{ceph_paper}, are yet to scale to the I/O performance required for large HPC deployments. The level of performance required for exascale class systems is demonstrated by the recent Frontier system deployment\cite{frontier_website}, where a 700PB, 5TB/s, Lustre filesystem is providing long term storage.

However, object stores have the potential to address long-standing performance and scalability issues associated with  POSIX filesystems for large-scale parallel I/O applications, such as applications using large numbers of small data files (KiBs up to a few MiBs) that can severely stress the metadata functionality of implementations such as Lustre and Spectrum Scale. Such issues can stem from constraints imposed by POSIX file system semantics, namely metadata prescriptiveness (per-file creation date, last-access date, permissions, etc.), and excessive consistency assurance\cite{nextplatform_lockwood}. Object storage technology can offer less restrictive semantics to POSIX file systems and can be unaffected by existing file system implementations or standards.

The Distributed Asynchronous Object Store (DAOS)\cite{daos-scfa2022} is a high-performance object store which features full user-space operation, with a RAFT-based consensus algorithm for distributed, transactional indexing, and byte-addressable access to NVM devices. It has demonstrated strong I/O performance as it comes to maturity, with high rankings in the I/O 500 benchmark\cite{io500} an indicator that is can scale to high metadata operation and I/O bandwidth rates. Previous research has demonstrated DAOS can provide high performance and resiliency for HPC applications\cite{daos-pdsw22}\cite{field_io_repo}\cite{daos_ecmwf_ipdps23}

However, the exploitation of object stores like DAOS for HPC I/O brings a number of usability and functionality questions that are yet to be fully explored. DAOS can be accessed in many different ways, and with a range of configurations, that bring a range of choices for both users and operators of such as system. In this work progress we discuss our investigation of the performance characteristics of the different access methodologies, and some of the configurations, that DAOS provides, and evaluate the best options for applications to utilise.

We present performance of standard I/O benchmarks that capture bulk I/O access patterns, using DAOS deployed on non-volatile storage hardware. A research HPC system containing nodes with 3 TiB of Intel's Optane Data Centre Persistent Memory Modules (DCPMMs) is used for benchmarking.




\section{DAOS}

Distributed Asynchronous Object Store (DAOS) is an open-source object store designed for massively distributed non-volatile data storage. Originally based on exploiting Intel's Optane memory, backed by NVMe storage devices, it has recently been expanded to support systems where Optane is not present, primarily by storing the metadata it holds in volatile memory backed by NVMe devices, rather than directly in Optane non-volatile memory. 

DAOS provides a low-level key-value storage interface on top of which other higher-level APIs, also provided by DAOS, are built, include a FUSE-based filesystem interface and custom interfaces for HDF5, MPI-I/O, and other common data storage formats/approaches. DAOS includes features such as transactional and non-blocking I/O, fine-grained I/O operations, end-to-end data integrity, and advanced data protection. The OpenFabrics Interfaces (OFI) library is used for low-latency communications over a wide range of high performance network back-ends. It also supports object classes that range from \texttt{S1} through to \texttt{SX} which enable objects to be distributed across DAOS engines in a similar manner to Lustre file striping.

DAOS enables a range of different methods for I/O from applications. The lowest level, requiring the most application adaption, is using key-value and array APIs (provided by \texttt{libdaos}) directly. There are also the options to use DFS (DAOS filesystem), either using \texttt{libdfs}, an API for file accesses and I/O on DAOS, or through \texttt{DFuse} (DAOS Fuse) which provides a user space filesystem mount point for the DAOS object store. libdfs and DFuse both allow applications to continue using file I/O whilst exploiting DAOS storage, whereas the DAOS API requires object store functionality to be added to applications.

\section{Methodology}

To assess the performance of DAOS, different I/O workloads have been generated and run in a research HPC system using the well-known IOR\cite{ior_repo} benchmark. 



For the tests in this analysis, the IOR benchmark has been run in both \textbf{easy} and \textbf{hard} modes, easy being a file-per-process, and hard being a single shared-file. IOR is invoked with a set of parameter values which instruct each client process to perform a set of I/O operations (transfers), to write and read its full data size. This is with the intent to assess the performance of a hypothetical optimised parallel application which is designed to minimise the number of I/O operations interacting with the storage and perform contiguous reads or writes.

IOR was originally designed to test performance of file systems, and therefore it can be used directly with its default POSIX back-end to test DAOS Fuse mounts. However, IOR also includes multiple additional back-ends that can be enabled to operate with the above-mentioned APIs, including libdaos' byte-array API, DFS, MPI-I/O, and HDF5. In this analysis, the DFS, MPI-I/O and HDF5 back-ends have been employed.

\section{Results}

The benchmarks we present have been conducted using the NEXTGenIO system\cite{ngio}, which is a research HPC system composed of dual-socket nodes with Intel Xeon Cascade Lake processors. Each socket has six 256 GiB first-generation Intel's Optane DCPMMs configured in AppDirect interleaved mode. We benchmarked using 8 server nodes with 2 DAOS engines per server node. 

Figures~\ref{fig:ior_easy} and~\ref{fig:ior_hard} present read and write bandwidth for the file-per-process and shared-file IOR test cases. We can see from the graphs that both the object class (i.e. S1, S2, or SX) and the access mechanisms (i.e. DAOS, HDF5, or MPI-I/O) impact the achieved performance.

\begin{figure}[htbp]
    \centering
    \begin{subfigure}[b]{190pt}
        \centering
        \includegraphics[width=190pt]{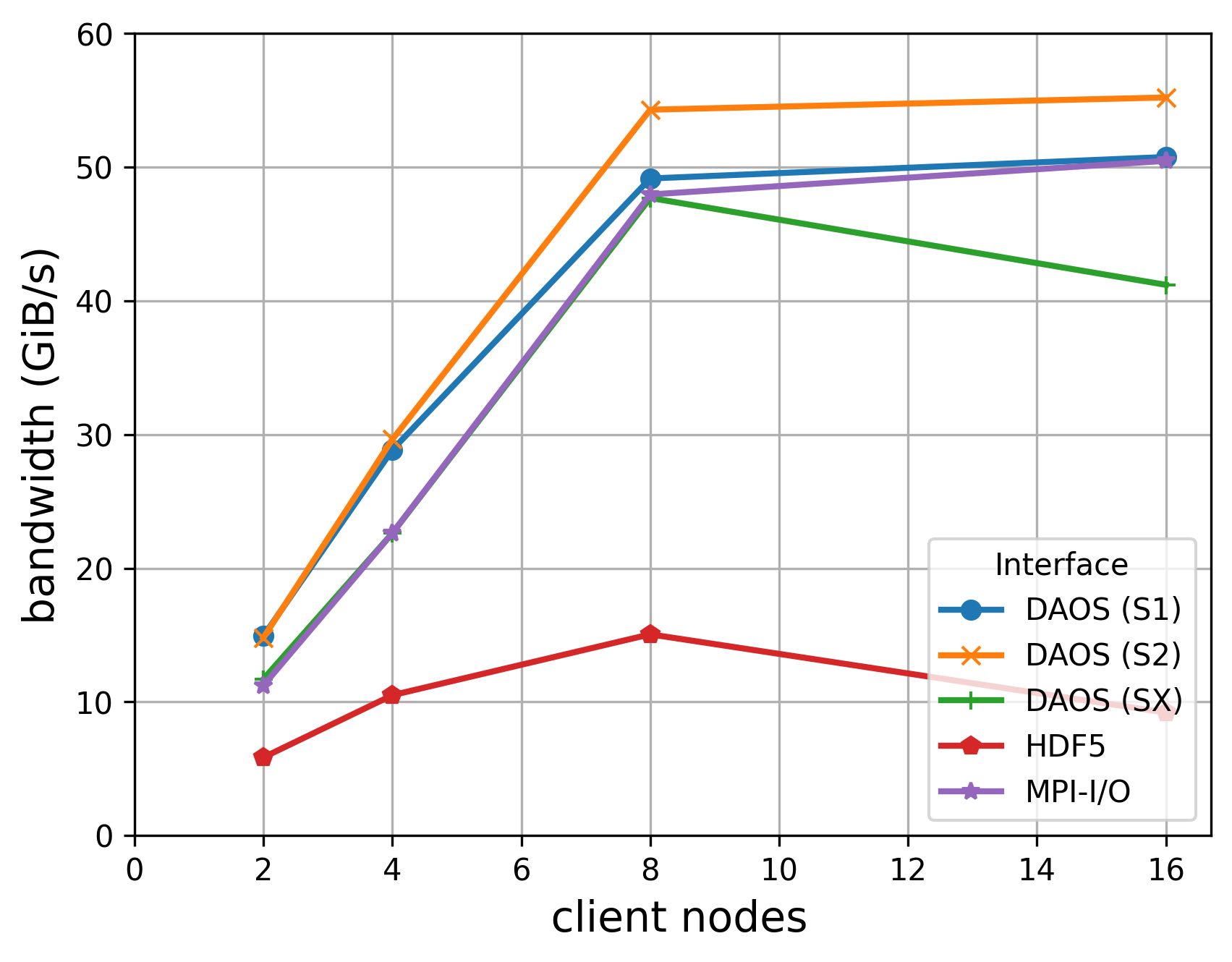}
        \caption{Read}
    \end{subfigure}
    \begin{subfigure}[b]{190pt}
        \centering
        \includegraphics[width=190pt]{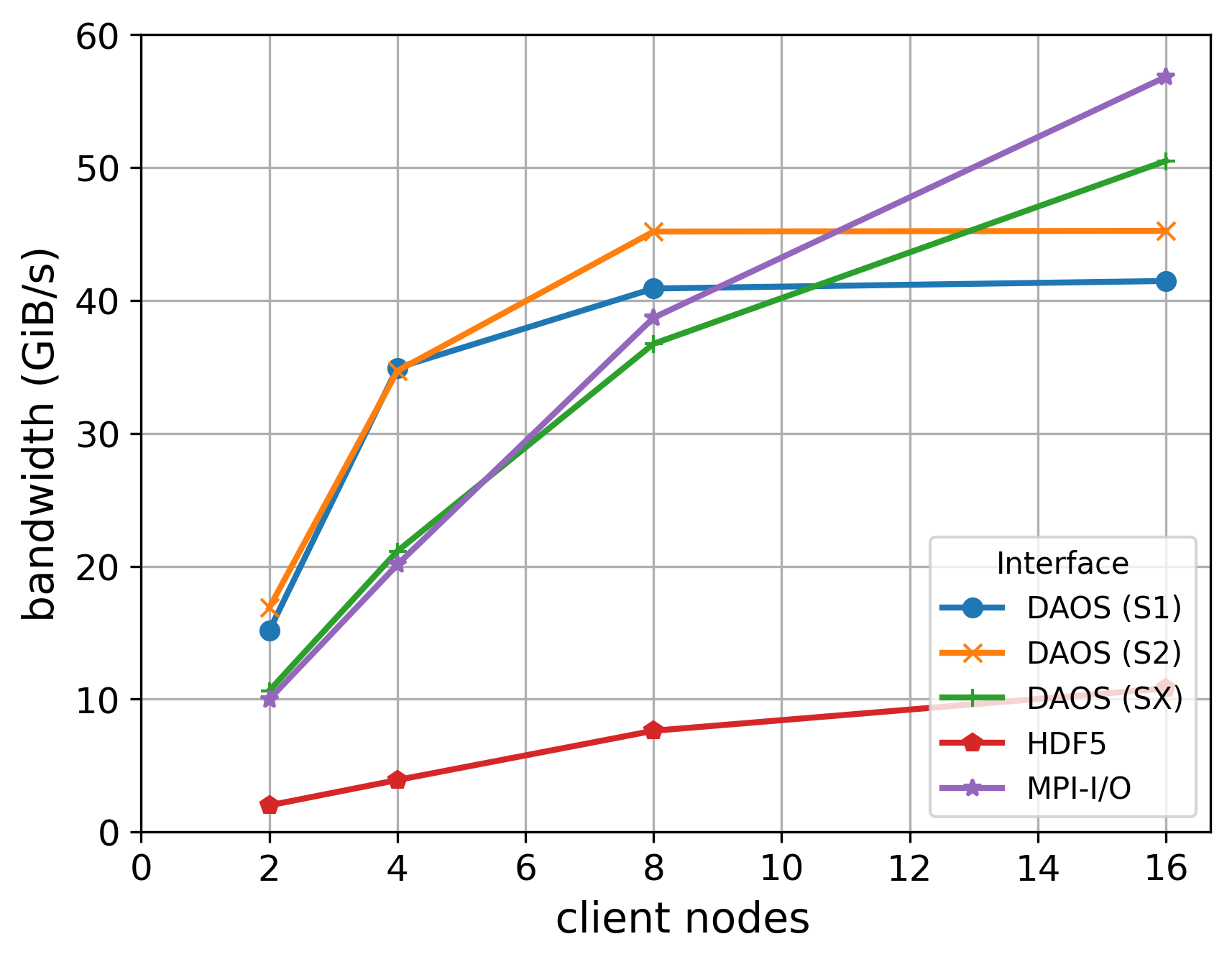}
        \caption{Write}
    \end{subfigure}
    \caption{IOR: File-per-process}
    \label{fig:ior_easy}
\end{figure}

It is clear that for the file-per-process scenario, having a small amount of object sharding (S2) gives the best performance for reading data, and good performance for writing until we scale up to the largest number of client nodes. Full sharding gives the best write performance for high contention scenarios (i.e. using the most client nodes) but lower performance for fewer writers. We can also see that using the DFS API (i.e. the DAOS lines) gives very similar performance to MPI-I/O using the DFuse mount, whereas HDF5 using the DFuse mount gives much lower performance, both for read and write.

\begin{figure}[htbp]
    \centering
    \begin{subfigure}[b]{190pt}
        \centering
        \includegraphics[width=190pt]{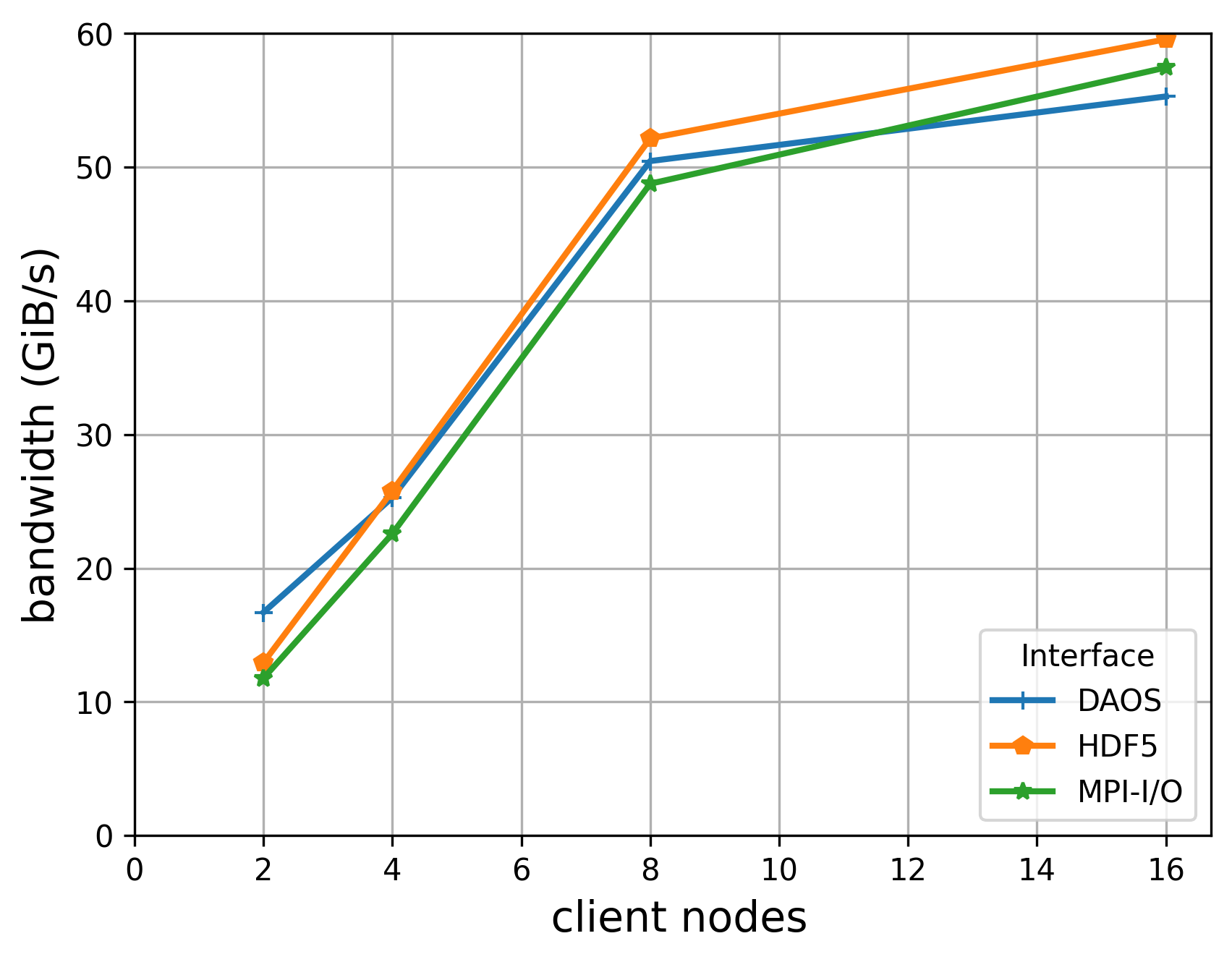}
        \caption{Read}
    \end{subfigure}
    \begin{subfigure}[b]{190pt}
        \centering
        \includegraphics[width=190pt]{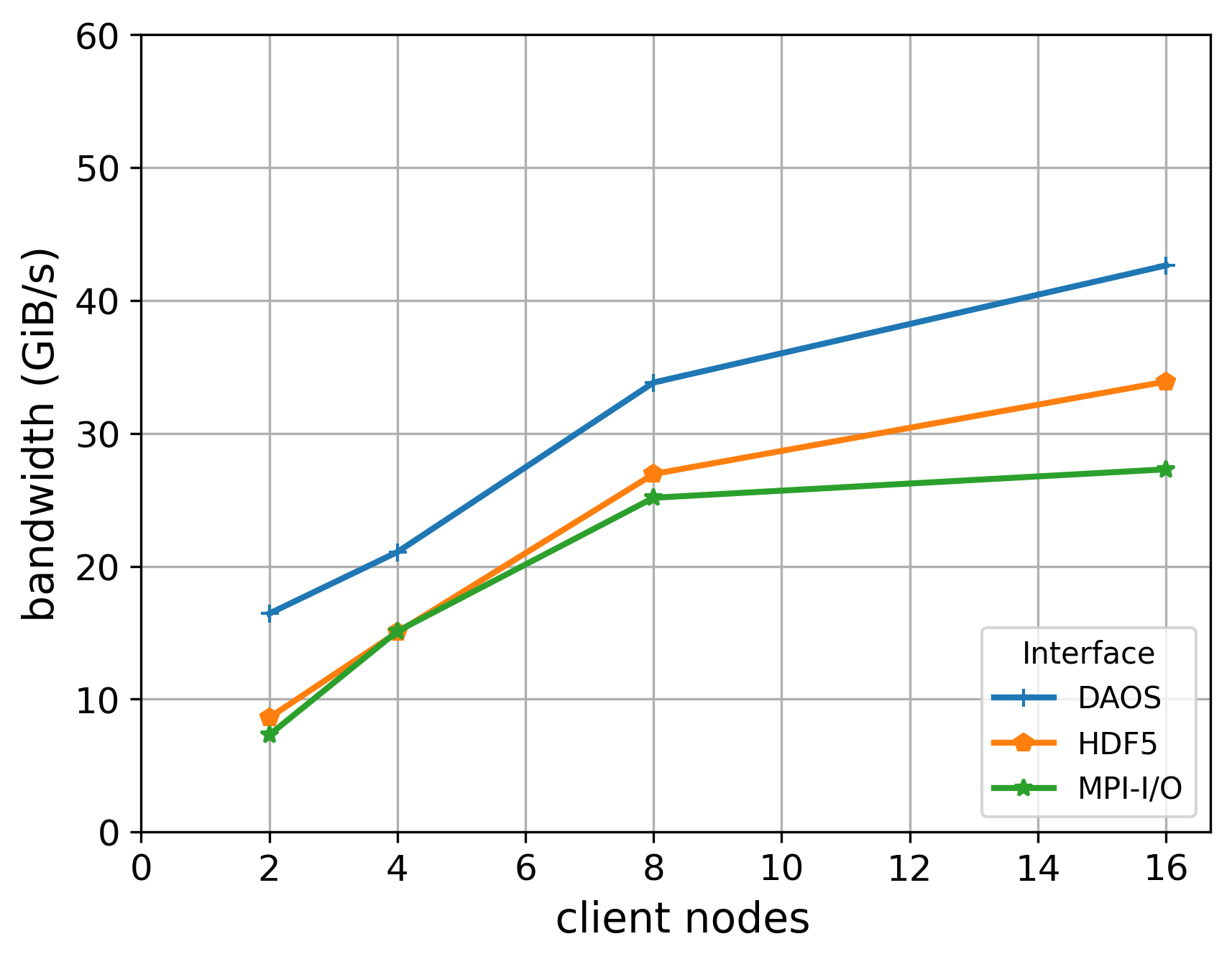}
        \caption{Write}
    \end{subfigure}
    \caption{IOR: Shared-file}
    \label{fig:ior_hard}
\end{figure}

The benchmarking done with a single shared file shows different performance characteristics, with similar performance achieved across interfaces. The DFS API gives the highest write bandwidth, and good read bandwidth, but both MPI-I/O and HDF5 using the DFuse mount also provide good read performance. Interesting, file-per-process and shared-file give similar overall performance, which is in stark contrast to the performance standard parallel filesystems provide.

\section{Conclusions}

We have presented some initial results from our current research to benchmark different access mechanisms and configurations for using DAOS from applications. We have demonstrated that both the object choice, and the interface/API used, can impact performance, but that exploiting DAOS via the available file APIs can still provide good performance, at least for bulk I/O, on a DAOS object store.

Future work will include extending benchmarking to use the DAOS API (rather than DFS or DFuse POSIX-based backends), and looking at some application specific I/O benchmarks to evaluate the kind of performance more varied usage patterns will experience.

\section*{Acknowledgment}

The work presented in this paper was carried out with funding by the European Union under the Destination Earth initiative (cost center DE3100, code 3320) and relates to tasks entrusted by the European Union to the European Centre for Medium-Range Weather Forecasts. Views and opinions expressed are those of the author(s) only and do not necessarily reflect those of the European Union or the European Commission. Neither the European Union nor the European Commission can be held responsible for them.

The NEXTGenIO system was funded by the European Union's Horizon 2020 Research and Innovation program under Grant Agreement no. 671951, and supported by EPCC, The University of Edinburgh.


\end{document}